\documentclass[a4paper,11pt]{article}
\usepackage{pos}

\usepackage[mathlines]{lineno}

\title{Understanding the background in dark matter searches by studying antinucleosynthesis in the laboratory with ALICE}
 \ShortTitle{Understanding the background in DM searches by studying antinucleosynthesis}

\manuallySeparateAuthors
\author*[a,b]{Sebastian Hornung}
\author{on behalf of the ALICE collaboration}

\affiliation[a]{Physikalisches Institut, Heidelberg University,\\
Im Neuenheimer Feld 226, 69120 Heidelberg, Germany}

\affiliation[b]{GSI Helmholtz Centre for Heavy Ion Research,\\
Planckstraße 1, 64291 Darmstadt, Germany}

\emailAdd{Sebastian.Hornung@cern.ch}

\abstract{
Antinuclei are considered to be one of the most promising probes in the indirect search for dark matter (DM) annihilation in space. However, in light of recent results on the production of light antinuclei in pp collisions at the LHC, an abundant production of light (anti)nuclei is also expected from Standard Model (SM) collisions of primary cosmic rays with the interstellar medium. Hence further precise measurements are required to constrain the production models of antinuclei in SM collisions to be sensitive to additional DM annihilation events.
 The most recent results of the ALICE collaboration on the production of antideuterons ($\overline{\mathrm{d}}$) and antihelium-3 ($^3\overline{\text{He}}$) in proton-proton (pp) and proton-lead (p--Pb) collisions are shown. These results provide valuable input for state-of-the-art calculations of the production models currently used to estimate the SM background to DM searches.
}

\FullConference{%
  40th International Conference on High Energy physics - ICHEP2020\\
  July 28 - August 6, 2020\\
  Prague, Czech Republic (virtual meeting)
}

\begin{document}
\maketitle
\section{Antinuclei flux in space}
\begin{figure}[hbt]
\centering
\includegraphics[width=0.45\textwidth]{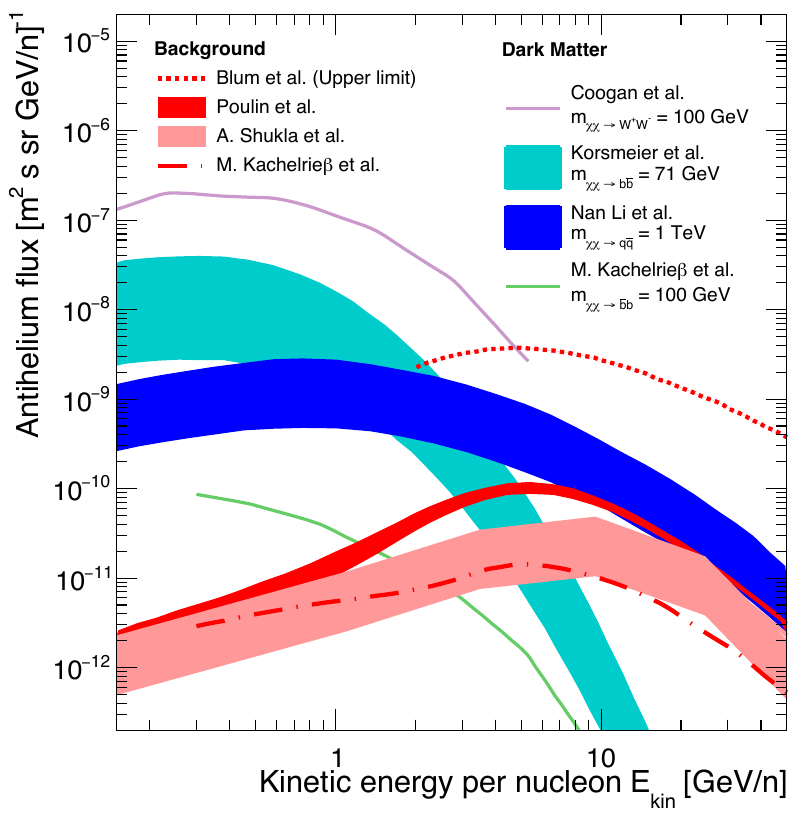}
\caption{$^3\overline{\text{He}}$ flux predictions from different DM models and astrophysical background calculations \cite{Coogan:2017pwt,Korsmeier:2017xzj,Li:2018dxj,Kachelriess:2020uoh,Blum:2017qnn, Poulin:2018wzu, Shukla:2020bql}. The error bands illustrate uncertainties in the coalescence momentum (see text for details), but also include propagation uncertainties. This figure is taken from \cite{vonDoetinchem:2020vbj}.}
\label{Fig:Predictions}
\end{figure}

Secondary antinuclei are expected to be produced via the decay or annihilation of dark matter particles in space \cite{vonDoetinchem:2020vbj}. After the propagation through space towards the earth, these particles could be observed as an excess of the light antinuclei flux above the expectations from interactions of primary cosmic rays (CR) and the interstellar medium (ISM) at low kinetic energies by satellite-born or balloon experiments.
As indicated in Figure \ref{Fig:Predictions}, which shows the predicted flux of $^3\overline{\text{He}}$ for different dark matter models and decays as well as different background calculations, the exact energy and strength of the excess related to dark matter depends on its characteristics \cite{vonDoetinchem:2020vbj}. Thus, it is important to be able to precisely predict the expected background from reactions of primary cosmic rays and the interstellar medium, which both consist mostly of protons and $^4\text{He}$ \cite{Shukla:2020bql,Gomez-Coral:2018yuk}.

Unfortunately, the production of light (anti)nuclei in hadronic interactions is not well understood and cannot be calculated from first principles. Thus, two classes of phenomenological models are employed to describe this process, the statistical hadronization and the coalescence models.
In the Statistical Hadronization Model (SHM) \cite{Andronic:2017pug}, the particles are produced from a medium in statistical equilibrium and their relative abundances are fixed at the chemical freeze-out, when the rate of inelastic scatterings between the constituents of the medium produced in the collision becomes negligible. Even though this model is very successful in describing the measured yield of light (anti)nuclei and even (anti)hypertriton in lead-lead collisions at the LHC using the grand canonical ensemble formulation \cite{Andronic:2017pug}, the description for smaller system, like in pp and p--Pb collisions, requires the exact conservation of charges, i.e. baryon number, strangeness number and electric charge.
Alternatively, the coalescence model is based on the assumption that (anti)nuclei are produced from (anti)nucleons if they are close enough in phase-space \cite{Kapusta:1980zz,Scheibl:1998tk}.
The key parameter of these models is the coalescence parameter $B_\mathrm{A}$, which is defined by 
\begin{linenomath}
\begin{align}
E_\mathrm{A}\frac{\mathrm{d}^{3}N_\mathrm{A}}{\mathrm{d}p_\mathrm{A}^{3}} = B_\mathrm{A} \left(E_{\mathrm{n}}\frac{\mathrm{d}^{3}N_{\mathrm{n}}}{\mathrm{d}p_{\mathrm{n}}^{3}}\right)^{A-Z} \left(E_{\mathrm{p}}\frac{\mathrm{d}^{3}N_{\mathrm{p}}}{\mathrm{d}p_{\mathrm{p}}^{3}}\right)^{Z} = B_\mathrm{A}{\left(E_{\mathrm{p}}\frac{\mathrm{d}^{3}N_{\mathrm{p}}}{\mathrm{d}p_{\mathrm{p}}^{3}}\right)^{A}} \Bigg|_{\vec{p}_{\mathrm{p}} = \vec{p}_\mathrm{A}/{A}}.
\end{align}
\end{linenomath}
It relates the invariant yield of nuclei with mass number $A$, $E_{A}(\mathrm{d}^{3}N_{A}/\mathrm{d}p_{A}^{3})$, to the one of its constituents.
At midrapidity and LHC energies, i.e. center-of-mass energies per nucleon-nucleon pair ($\sqrt{s_\mathrm{NN}}$) in the order of a few TeV, the invariant yield of protons, $E_{\mathrm{p}}(\mathrm{d}^{3}N_{\mathrm{p}}/\mathrm{d}p_{\mathrm{p}}^{3})$, is expected to be identical to the one of neutrons, and it is therefore used to describe both protons and neutrons. The momentum of the proton, $\vec{p}_\mathrm{p}$, is given by the one of the nuclei, $\vec{p}_A$, divided by their mass number.
The coalescence parameter can also be related to the production probability of the nucleus and can be calculated from the overlap of the wave function of the nucleus with the phase space distribution of its constituent nucleons via the Wigner formalism \cite{Bellini:2018epz}.
The estimate of the background from the interactions of CRs and the ISM is usually based on the coalescence approach employing either the analytical coalescence model \cite{Korsmeier:2017xzj, Poulin:2018wzu, Shukla:2020bql} or a model based on the scaling of $B_A$ with the radius of the emission source \cite{Blum:2017qnn,Bellini:2018epz}.
In the former, all  (anti)nucleons with momenta within a sphere with a radius $p_0$, the coalescence momentum,  form a bound state. In more recent implementations of this model, the (anti)nucleons are produced employing an hadronic event generator to take into account their possible correlations \cite{Korsmeier:2017xzj,Shukla:2020bql} and the coalescence process is taken into account by an afterburner. The coalescence momentum cannot be calculated within these models and has to be extracted from measurements. In \cite{Shukla:2020bql} also the observed kinetic energy dependence of $p_0$ has been taken into account.
Both model descriptions strongly rely on measurements of the production yields of light antinuclei in pp and pA collisions, as for example performed by the ALICE collaboration. The estimated antinuclei flux strongly depends on the precise knowledge of the production rate together with a good understanding of the propagation processes of the antinuclei through space.
In this proceeding, we focus on the application of the measurements of antideuterons and $^3\overline{\text{He}}$ in pp and p--Pb collisions at different collision energies provided by the ALICE collaboration in the context of the predictions of the expected top-of-the-atmosphere of antinuclei. 

\section{Production of light (anti)nuclei in ion collisions}

\begin{figure}[hbt]
\centering
\includegraphics[width=0.34\textwidth]{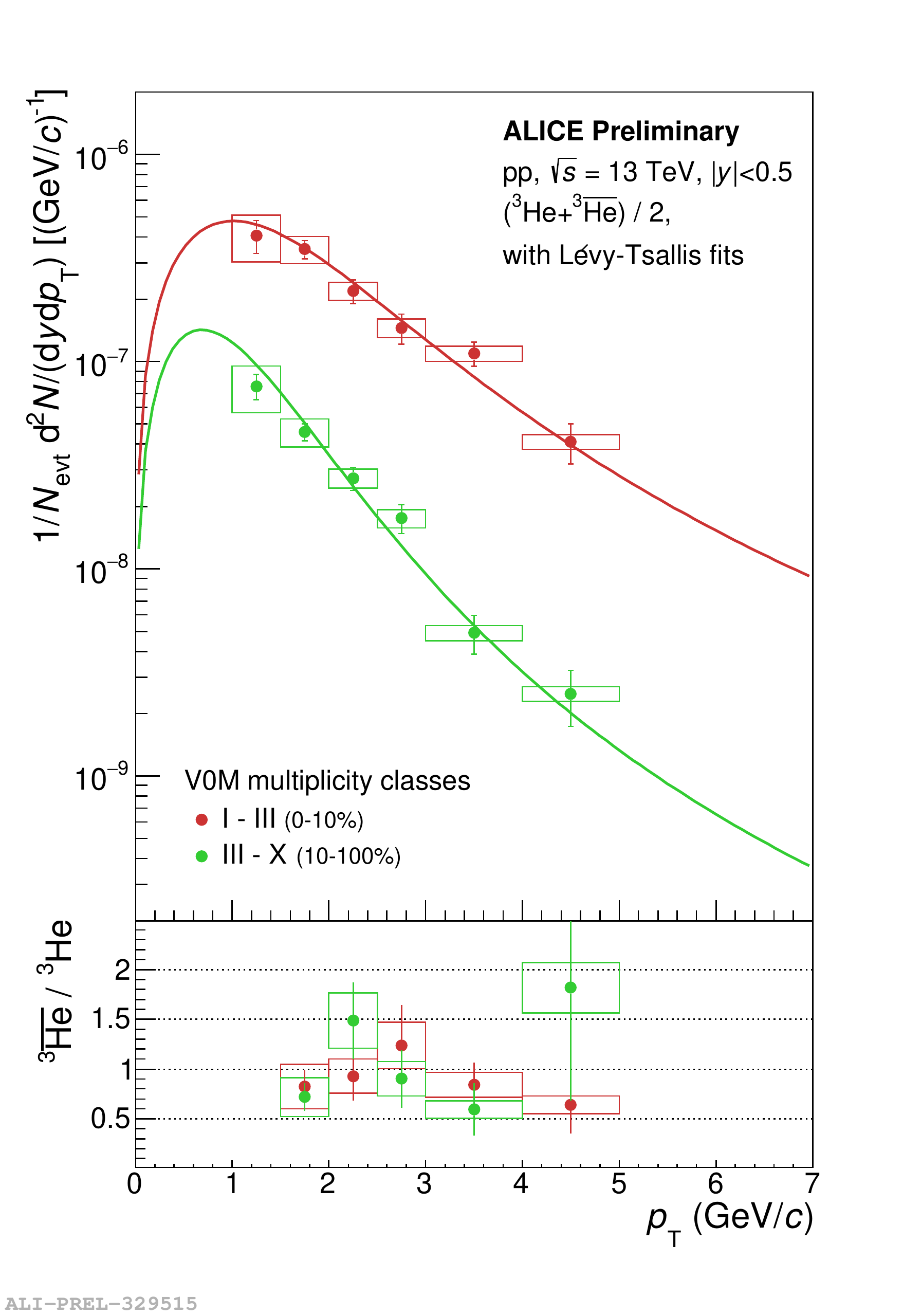}
\includegraphics[width=0.35\textwidth]{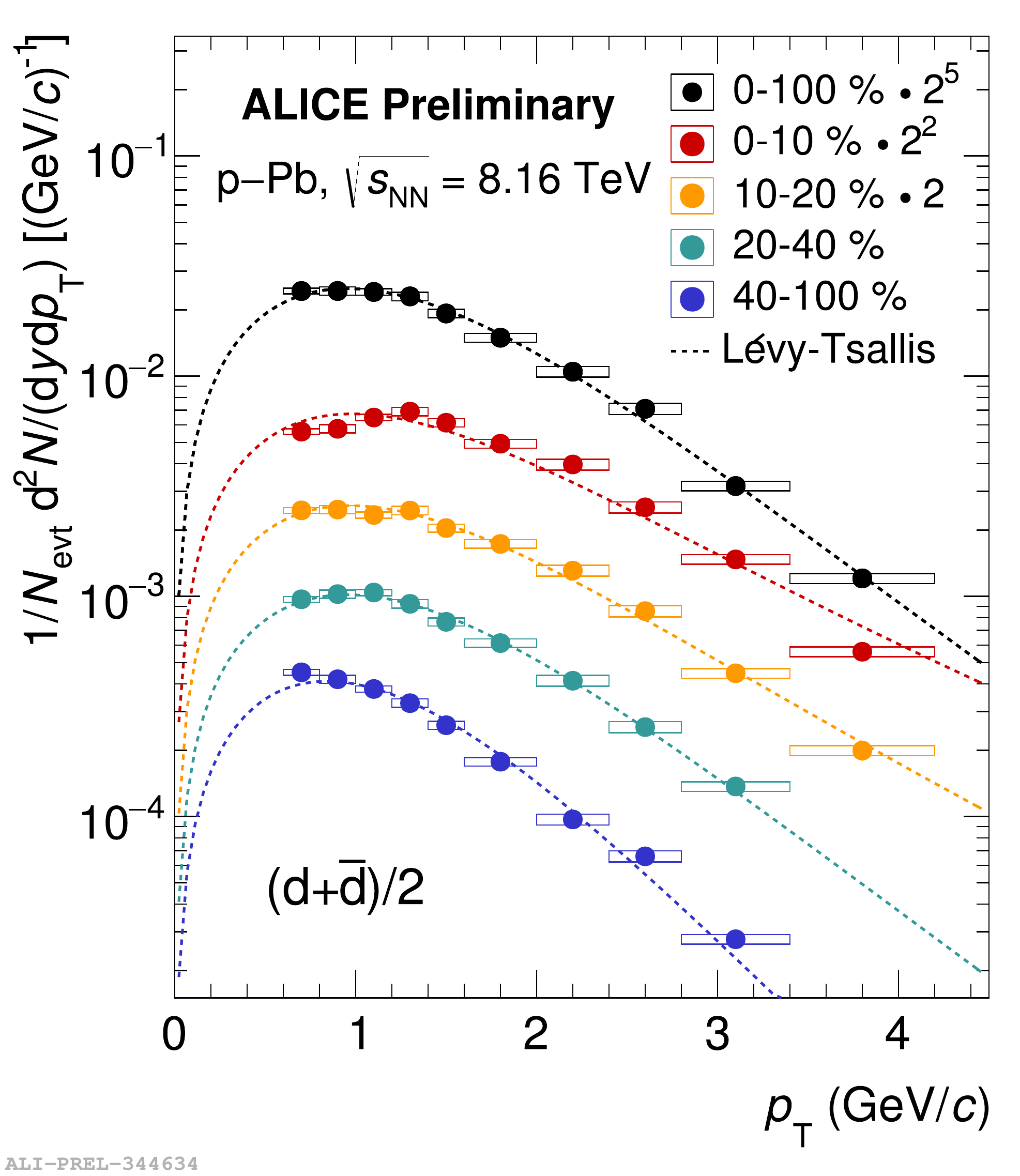}
\caption{$p_\mathrm{T}$ spectra of (anti-)$^3\text{He}$ measured in pp collisions at $\sqrt{s} = 13$~TeV (left) and (anti)deuterons measured in p--Pb collisions at $\sqrt{s_\mathrm{NN}} = 8.16$~TeV (right) in different multiplicity classes.
}
\label{Fig:Spectra}
\end{figure}

The current estimates for top-of-the-atmosphere flux of secondary antinuclei from the interaction of primary cosmic rays and the interstellar medium are based on measurements of the production of light antinuclei in pA and pp collisions including the published results from ALICE~\cite{Acharya:2017fvb,Adam:2015vda}. Recently, the ALICE collaboration has published new results for the transverse-momentum spectra of (anti)deuterons in pp collisions at $\sqrt{s} = 13$~TeV \cite{Acharya:2020sfy} and (anti-)$^3\text{He}$ in p--Pb collisions at $\sqrt{s_\mathrm{NN}} = 5.02$ \cite{Acharya:2019xmu}. In addition, the preliminary results for (anti-)$^3\text{He}$ in pp collisions at $\sqrt{s} = 13$~TeV and (anti)deuterons in p--Pb collisions at $\sqrt{s_\mathrm{NN}} = 8.16$~TeV are summarized in Figure \ref{Fig:Spectra}.

These preliminary measurements have not yet been used for the current predictions in \cite{Shukla:2020bql}, but will help to reduce the differences between the predictions and their respective uncertainty bands. These measurements can be exploited to directly compare the corresponding predictions from the models used to estimate the background in DM searches, as done in \cite{Shukla:2020bql}. Moreover, the corresponding coalescence parameter $B_\mathrm{A}$ can be either used to extract the coalescence momentum $p_0$ or to compare with the models based on the scaling with the size of the emission region \cite{Blum:2017qnn}.

\section{Coalescence parameter}
\begin{figure}[hbt]
\centering
\includegraphics[width=0.48\textwidth]{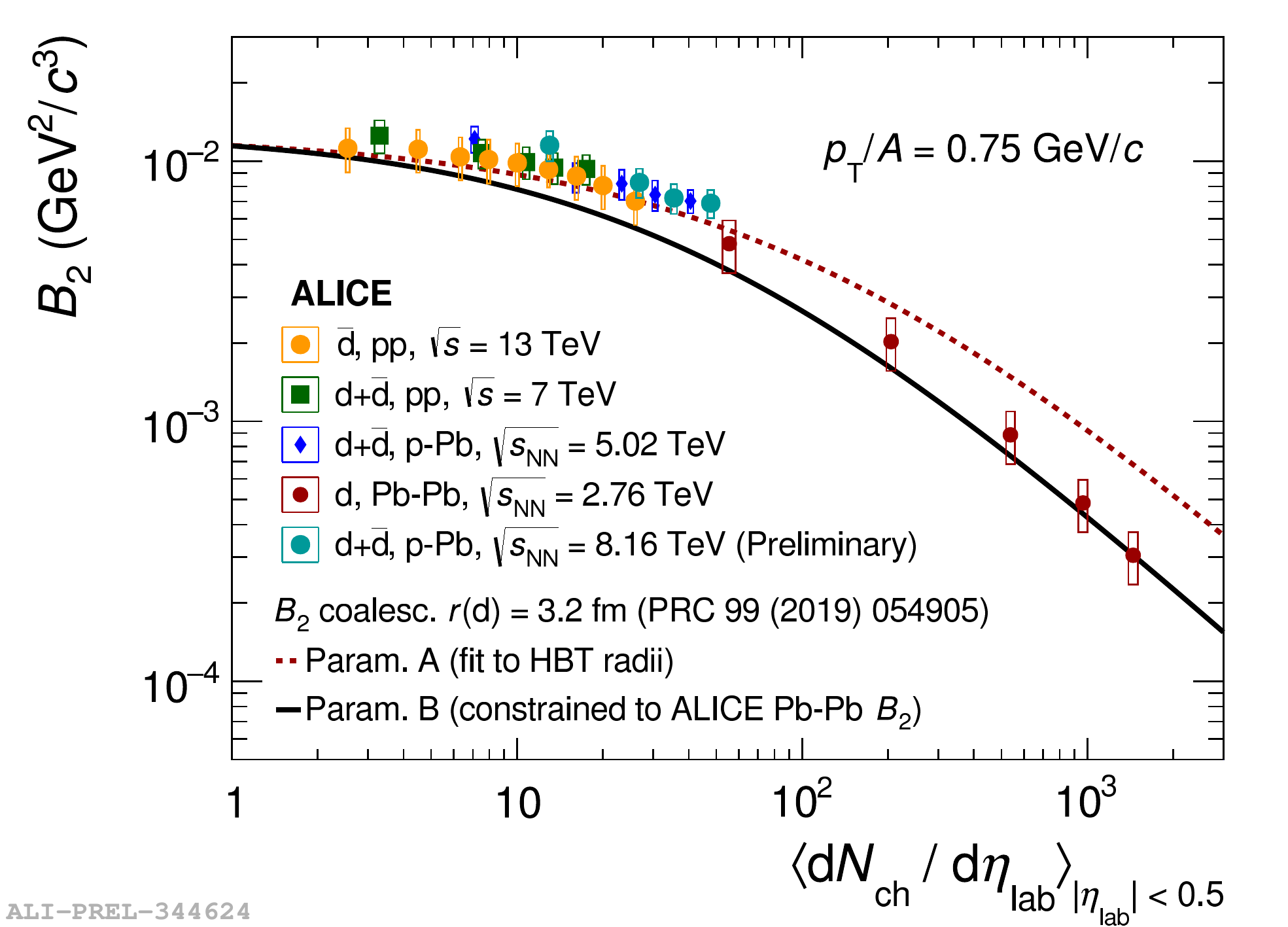}
\includegraphics[width=0.45\textwidth]{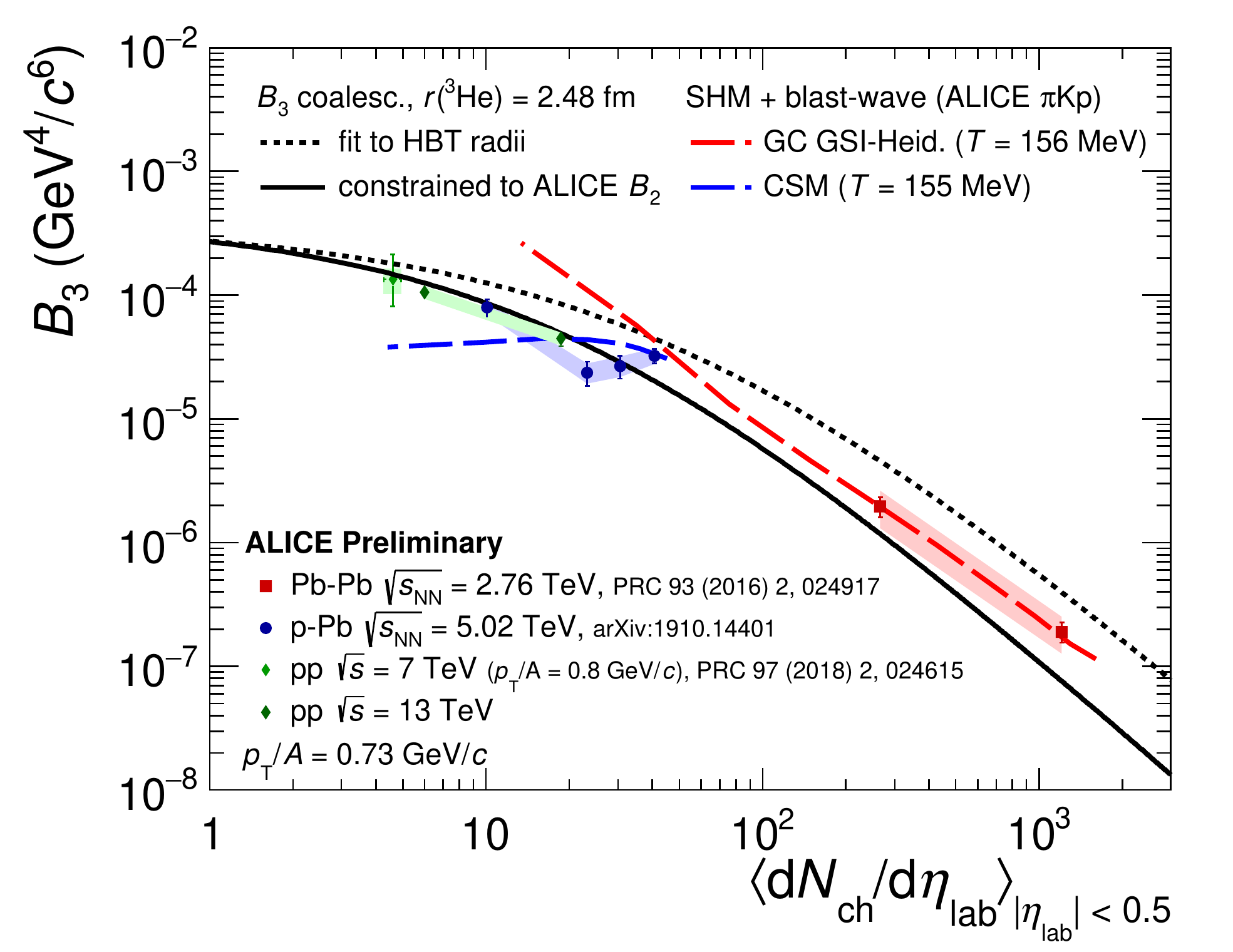}
\caption{$B_2$ and $B_3$ measured as a function of the mean charged-particle multiplicity density at midrapidity are compared to the coalescence model expectations from \cite{Bellini:2018epz} at a fixed value of $p_\mathrm{T}/A = 0.75$ GeV/$c$ and $p_\mathrm{T}/A~=~0.73$~GeV/$c$, respectively. Two different parameterizations for the radius of the emission region as a function of $\langle\text{d}N_\text{ch}/\text{d}\eta_\mathrm{lab}\rangle_{|\eta_\mathrm{lab}|<0.5}$ are used. In the right panel, the expectations for the grand canonical (GC) \cite{Andronic:2017pug} and the canonical statistical hadronization model (CSM) \cite{Vovchenko:2018fiy} are also indicated.}
\label{Fig:B2and3vsMult}
\end{figure}
The evolution of $B_A$ with the size of the emission region can be studied by measuring $B_A$ as a function of $p_\mathrm{T}/A$ for several collision systems and energies \cite{Acharya:2019rgc, Acharya:2019rys, Adam:2015vda,Acharya:2019xmu}.
For a fixed value of $p_\mathrm{T}/A = 0.75$ GeV/$c$ and $p_\mathrm{T}/A = 0.73$ GeV/$c$, $B_2$ and $B_3$ are shown as a function of the mean charged-particle multiplicity density at midrapidity, $\langle\text{d}N_\text{ch}/\text{d}\eta_\mathrm{lab}\rangle_{|\eta_\mathrm{lab}|<0.5}$, in Figure \ref{Fig:B2and3vsMult}.
The mean charge-particle multiplicity density is proportional to the volume of the emission region of the (anti)nuclei,
$
R \propto \left(\langle \mathrm{d}N_\mathrm{ch}/d\eta_\mathrm{lab}\rangle_{|\eta_\mathrm{lab}| < 0.5} \right)^{1/3}
$.
Thus, these measurements directly prove the scaling of the coalescence parameter with the size of the emission region as employed by one of the coalescence models used to estimate the flux of secondary antinuclei at the top of the atmosphere~\cite{Blum:2017qnn}.
The measurements show a smooth transition from low charged-particle multiplicity densities, which refer to a small emission region, to large ones. This indicates that the production mechanism in small collision systems evolves continuously to the one in large systems and also that a single mechanism sensitive to the size of the emission region could be possible.
The experimental results are compared to calculations obtained from the coalescence approach \cite{Bellini:2018epz} for two different parameterizations for the size of the emission region as a function of $\langle\text{d}N_\text{ch}/\text{d}\eta_\mathrm{lab}\rangle_{|\eta_\mathrm{lab}|<0.5}$. The theoretical calculations agree with the trend observed in data.

\section{Summary and outlook}
The ALICE collaboration has measured the production of light (anti)nuclei in different collision systems and at different energies providing crucial input information for the coalescence models used to estimate the expected background from interactions of primary cosmic ray particle  and the interstellar medium. In particular, the most recent measurements of the production of antideuterons and $^3\overline{\text{He}}$ can be used to validate and constrain the model predictions.
The theoretical descriptions based on the scaling of the coalescence parameters $B_2$ and $B_3$ with the size of the emission region are in agreement with the observed smooth evolution of the coalescence parameters as a function of $\langle\text{d}N_\text{ch}/\text{d}\eta_\mathrm{lab}\rangle_{|\eta_\mathrm{lab}|<0.5}$, which is proportional to the volume of the emitting source.

In future data taking campaigns, the ALICE collaboration will be able to provide more precise and more differential results on the production of antideuterons and $^3\overline{\text{He}}$ as well as first measurements of the production of $^4\overline{\text{He}}$ in pp and p--Pb collisions at the LHC. These results will help to further constrain the predictions of the expected background interactions of CR and the ISM which affects the searches for an excess of the  flux of antinuclei at the top of the atmosphere.

\bibliographystyle{JHEP}
\bibliography{NucleiBiblio}

\end{document}